\documentclass[reqno, 12pt]{amsart}
\usepackage[ansinew]{inputenc}
\usepackage{amsfonts}
\usepackage{latexsym}
\usepackage{amsmath}
\usepackage{graphicx}
\usepackage{color}

\newtheorem{defin}{Definition}[section]

\begin{document}

\footnote{*Corresponding author: Jian-Guo Liu, Email: 20101059@jxutcm.edu.cn; Wen-Hui Zhu, Email: 415422402@qq.com.}

\begin{center}{\bf \large Interaction phenomenon for variable coefficient Kadomtsev-Petviashvili equation by  utilizing variable coefficient bilinear neural network method}
\end{center}

\vskip 4mm

\begin{center} Jian-Guo Liu$^{1*}$, Wen-Hui Zhu$^{2*}$
\end{center} \vskip 2mm

\noindent$^{1}$ College of Computer, Jiangxi University of
Chinese Medicine, Jiangxi 330004, China;\\
\noindent$^{2}$ Institute of artificial intelligence, Nanchang
Institute of Science and Technology, Jiangxi 330108, China;
 \vskip 6mm

{\bf Abstract:}\,In this paper,  a variable coefficient Bilinear neural network method is proposed to deal with the analytical solutions of variable coefficient nonlinear partial differential equations.  As an example, a Kadomstev-Petviashvili equation with variable coefficients is investigated by using the  variable coefficient Bilinear neural network method.  By establishing ``3-2-2-1'' and ``3-3-2-1'' models respectively, rich analytical solutions of the variable coefficient Kadomstev-Petviashvili equation are obtained. By choosing some special values of the parameters, the dynamics properties are demonstrated in some three-dimensional and density graphics.  \vskip 2mm
 {\bf Keywords:}\,
variable coefficient Bilinear neural network method, variable coefficient Kadomstev-Petviashvili equation, analytical solutions, dynamic properties. \vskip
2mm

{\bf 2010 Mathematics Subject Classification: 35Q51, 35G99, 33F10 }
\vskip 4mm

\noindent {\bf 1. Introduction}\\

\quad  Partial differential equations (PDEs) have a wide range of applications, from the simulation of seismic waves on Earth to the spread of infectious diseases through human populations. Engineers, scientists, and mathematicians have resorted to PDEs to describe complex phenomena involving many independent variables [1,2]. However, the process of solving PDEs is extremely difficult, traditional numerical methods are very complex and require a lot of computation. Researchers have proved that it can obtain approximate solutions faster than traditional PDE solvers through neural network experiments [3], such as Convolutional Neural Networks [4], Recurrent Neural Network [5], Generative Adversarial Networks [6], DeepONet [7], physic informed neural network (PINN) [8-10] and  so on. Recently, Zhang [11] et al proposed the Bilinear Neural Network Method (BNNM) for investigating the analytical solutions of PDEs based on the bilinear method and the Neural Network model.  This method was  used in many constant coefficient PDEs, such as p-generalized B-type Kadomstev-Petviashvili (KP) equation [12], Caudrey-Dodd-Gibbon-Kotera-Sawada-like equation [13], (3+1)-dimensional Jimbo-Miwa equation [14], Boiti-Leon-Manna-Pempinelli equation [15] and so on. However, the BNNM has never been used in variable coefficient PDEs. Prompted by this, we modify this method and apply it to study the analytical solutions of the KP equation with variable coefficients.

\quad In this article,  under consideration is the following  KP equation with variable coefficients [16]
\begin{eqnarray}
\alpha  u_x^2+\alpha  u u_{xx}+\beta  u_{xxxx}-\gamma u_{yy}+u_{xt}=0,
\end{eqnarray}where $u=u(x,y,t)$.  $\alpha=\alpha (t)$, $\beta=\beta (t)$ and $\gamma=\gamma (t)$ mean the coefficients
of nonlinearity, dispersion and disturbed wave velocity, respectively. Eq. (1) represents waves in ferromagnetic media,
water waves of long wavelength with weakly non-linear
restoring forces and frequency dispersion. Wu et al. [16-18] have studied the
B\"{a}cklund transformation, soliton, Wronskian and
Gramian solutions for Eq. (1). Jia [19] et al. have investigated the lump and
rogue waves of Eq. (1). The lump, breather wave and interaction solutions have been obtained in Ref. [20,21]. Multiple rogue wave solutions were derived based on the variable-coefficient symbolic computation approach [22].  Nonautonomous lump solutions were presented in Ref. [23].  Double-periodic soliton and non-traveling wave solutions  have been given in Ref. [24]. Solitary and lump waves interaction were investigated by using a modified Ans\"{a}tz with variable coefficients [25]. However, the bilinear neural network method has not been used in Eq. (1), which will become
our main work.

\quad  Let's do the following transformation{\begin{eqnarray}
\alpha=\frac{6 \beta }{\Theta _0},\,\,\,\,\,\, u=2\,\Theta _0\,(ln\xi)_{xx}, \,\,\,\xi=\xi(x,y,t),
\end{eqnarray}
}Eq. (1) has the
bilinear form\begin{eqnarray}
&&(\beta D_x^4-\gamma D_y^2+ D_t D_x) \xi\cdot \xi=\xi  (\beta  \xi_{xxxx}-\gamma  \xi_{yy}+\xi_{xt})\nonumber\\&&+3 \beta  \xi_{xx}^2-4 \beta  \xi_x \xi_{xxx}+\gamma  \xi_y^2-\xi_t \xi_x=0,
\end{eqnarray}where $\Theta _0$ is an arbitrary constant.

\quad The paper is organized as follows. Section 2
proposes a variable coefficient Bilinear neural network method (vcBNNM); Section 3
obtains the analytical solutions of the KP equation with variable coefficients by using the vcBNNM. Meanwhile, the dynamic properties are described in some three-dimensional and contour plots. Section 4
 makes a summary.\\

\noindent {\bf 2. vcBNNM}\\

\quad  In order to apply the BNNM in Ref. [11] to the variable coefficient PDEs, we will improve the BNNM as follows (named vcBNNM)\\
{\bf (I)}\,  The tensor formula of vcBNNM is supposed as follows
\begin{eqnarray}
\xi=\omega_{l_n,\xi}\,F_{l_n}(\xi_{l_n}),
\end{eqnarray}where $\xi=\xi(x, y, \cdots , t)$, $\xi_{l_n}=\xi_{l_n}(x, y, \cdots , t)$, $\omega_{l_n,\xi}$ is the weight coefficient of neuron $l_n$ to
$\xi$. $F$ is a generalized activation function that can be defined arbitrarily and must
be satisfied that $F_{l_n}(\xi_{l_n})\geq 0$ in the last layer. $l_n={m_{n-1} + 1, m_{n-1}+
2, ... , n}$ denotes the nth layer space of the neural
network model. $\xi_{l_i}$ is assumed as follows{\begin{eqnarray}
\xi_{l_i}=\omega_{l_{i-1},l_i}\,F_{l_{i-1}}(\xi_{l_{i-1}})
+\omega_{l_i}(t),\,\,\,i=1, 2, \cdots , n,
\end{eqnarray}}where $l_0={x, y, \cdots , \omega_{l_0}(t)}$, $l_1={1, 2, \cdots , m_1}$, $l_i={m_{i-1} + 1, m_{i-1}+
2, ... , m_i} (i=2, 3, \cdots , n-1)$, $\omega_{l_i}(t)$  represents the  threshold, which is different from the constant hypothesis in Ref. [11]. This neural network tensor model is shown in Fig. 1.\\

{\bf (II)} Substituting Eq. (4) into the Hirota bilinear form of variable coefficient PDEs and
making the coefficient of $F_{l_i}(\xi_{l_i})$ and its derivatives of all orders equal to zero, a sets of algebraic equations will be obtained. With the aid of Mathematica,  the values of the parameters in  Eq. (4) and Eq. (5) can be derived. Finally, substituting these
values and nonlinear neural network tensor for-
mula Eq. (10) into Eq. (4) and Eq. (5), the
corresponding analytical solutions for the variable coefficient PDEs can be
presented.

\includegraphics[scale=0.55,bb=-120 210 10 10]{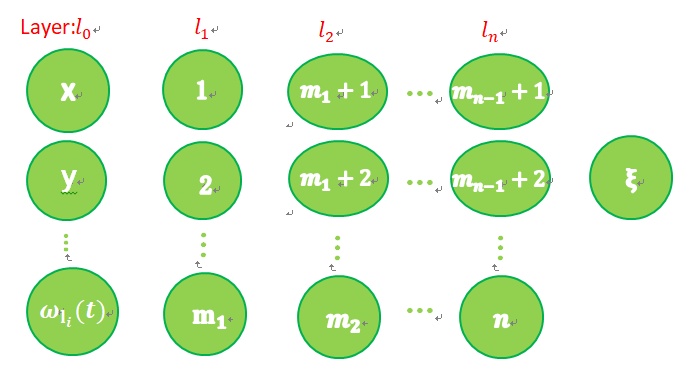}
\vspace{4.2cm}
\begin{tabbing}
\textbf{Fig. 1}. Nonlinear neural network of Eq. (4) and Eq. (5).
\end{tabbing}

\includegraphics[scale=0.55,bb=20 320 10 10]{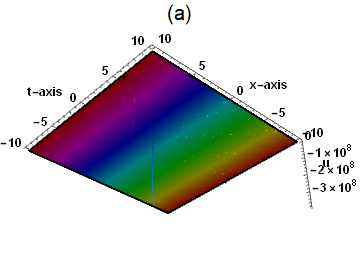}
\includegraphics[scale=0.43,bb=-480 390 10 10]{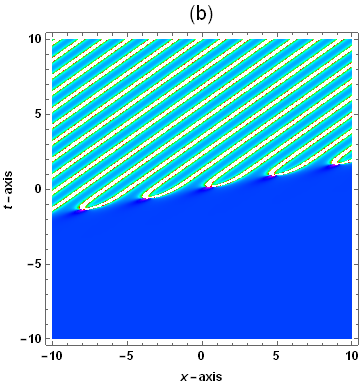}
\vspace{5.8cm}
\begin{tabbing}
\textbf{Fig. 2}. $ \eta_1=\omega _{1,2}=\omega _{2,1}=\omega _{2,3}=\Theta_0=1$, $\eta_2=\omega _{1,1}=\omega _{1,3}=2$, $\omega _{2,2}=-2$,\\ $y=0$, (a) three-dimensional plot,  (b) density plot.
\end{tabbing}

\noindent {\bf 3. Application}\\

\quad To verify the effectiveness of the vcBNNM in Section 2, we use the variable coefficient KP equation (1) as an example and select a ``3-2-2-1''
variable coefficient neural network model, which represents that the input layer $l_0=\{x, y, \omega_{l_i}(t)\} (i=1, 2)$ has 3 neurons, the hidden layer $l_1=\{1, 2\}$ has 2 neurons, the hidden layer $l_2=\{3, 4\}$ has 2 neurons and the print layer $\xi$ has 1 neuron. According to the idea of vcBNNM, we suppose
{\begin{eqnarray} &&\xi_1= \omega _{3,1}(t)+x \omega _{1,1}+y \omega _{2,1},\nonumber\\
&&\xi_2=\omega _{3,2}(t)+x \omega _{1,2}+y \omega _{2,2},\nonumber\\
&&\xi_3=\omega _{1,3} F_1\left(\xi _1\right)+\omega _{2,3}
   F_2\left(\xi _2\right)+\omega _{3,3}(t),\nonumber\\
&&\xi_4=\omega _{1,4} F_1\left(\xi _1\right)+\omega _{2,4}
   F_2\left(\xi _2\right)+\omega _{3,4}(t),\nonumber\\
&&\xi=\omega _{1,5} F_3\left(\xi _3\right)+\omega _{2,5}
   F_4\left(\xi _4\right)+\omega _{3,5}(t),
\end{eqnarray}}where $F_1(\xi_1)=\cos \xi_1$, $F_2(\xi_2)=\exp \xi_2$, $F_3(\xi_3)=\xi_3^2$, $F_4(\xi_4)=\xi_4^2$, $\omega _{i,j} (i=1, 2; j=1, 2, 3, 4, 5)$ is unknown constant, $\omega _{3,j}(t) (j=1, 2, 3, 4, 5)$ is unknown function, which is different from the constant hypothesis in Ref. [11], and the purpose is to deal with the variable coefficients in Eq. (1). This is also the biggest difference and innovation between the vcBNNM and the BNNM. Substituting Eq. (6) into Eq. (3) by using Mathematica software, we have
{\begin{eqnarray} &&\omega _{2,4}=\frac{\omega _{1,4} \omega _{2,3}}{\omega _{1,3}}, \omega _{2,5}=-\frac{\omega _{1,3}^2 \omega _{1,5}}{\omega _{1,4}^2}, \omega _{3,1}(t)=\eta _2+\nonumber\\ &&\frac{\left(-\omega _{2,2}^2 \omega _{1,1}^4+2 \omega _{1,2} \omega
   _{2,1} \omega _{2,2} \omega _{1,1}^3+\omega _{1,2}^2 \left(2 \omega
   _{1,1}^2+3 \omega _{1,2}^2\right) \omega _{2,1}^2\right) \int
   \gamma  \, dt}{3 \omega _{1,1} \omega _{1,2}^2 \left(\omega
   _{1,1}^2+\omega _{1,2}^2\right)},\nonumber\\ && \omega _{3,2}(t)=\eta _1+[[\omega _{1,2}^2 \left(\omega _{1,2}^2-6 \omega
   _{1,1}^2\right) \omega _{2,1}^2-2 \omega _{1,1} \omega _{1,2}
   \left(\omega _{1,2}^2-6 \omega _{1,1}^2\right) \omega _{2,2} \omega
   _{2,1}\nonumber\\ &&+\omega _{1,1}^2 \left(4 \omega _{1,2}^2-3 \omega
   _{1,1}^2\right) \omega _{2,2}^2] \int \gamma \, dt]/[3
   \omega _{1,1}^2 \omega _{1,2} \left(\omega _{1,1}^2+\omega
   _{1,2}^2\right)],\nonumber\\ && \omega _{3,5}(t)=\omega _{1,5} [\frac{\omega _{1,3}^2 \omega _{3,4}(t){}^2}{\omega
   _{1,4}^2}-\omega _{3,3}(t){}^2], \beta=-\frac{\left(\omega _{1,2} \omega _{2,1}-\omega _{1,1} \omega
   _{2,2}\right){}^2 \gamma }{3 \omega _{1,1}^2 \omega _{1,2}^2
   \left(\omega _{1,1}^2+\omega _{1,2}^2\right)},
\end{eqnarray}}where $\eta _1$ and $\eta _2$ are integration constants. Substituting Eq. (6) and Eq. (7) into Eq. (2), the corresponding analytical solution for Eq. (1) can be obtained as follows
{\begin{eqnarray}  &&u=[2 \Theta _0 [\left(\omega _{1,5} F_3\left(\xi_3\right)+\omega
   _{2,5} F_4\left(\xi_4\right)+\omega _{3,5}(t)\right) [\omega _{1,5}
   F_3''\left(\xi_3\right) [\omega _{1,1} \omega _{1,3} F_1'\left(\xi_1\right)\nonumber\\&&+\omega _{1,2}
   \omega _{2,3} F_2'\left(\xi_2\right)]{}^2+\omega _{1,5} F_3'\left(\xi_3\right) [\omega
   _{1,3} F_1''\left(\xi_1\right) \omega _{1,1}^2+\omega _{1,2}^2 \omega _{2,3}
   F_2''\left(\xi_2\right)]\nonumber\\&&+\omega _{2,5} F_4'\left(\xi_4\right) [\omega
   _{1,4} F_1''\left(\xi_1\right) \omega _{1,1}^2+\omega _{1,2}^2 \omega _{2,4}
   F_2''\left(\xi_2\right)]+\omega _{2,5} [\omega _{1,1} \omega
   _{1,4} F_1'\left(\xi_1\right)\nonumber\\&&+\omega _{1,2} \omega _{2,4} F_2'\left(\xi_2\right)]{}^2
   F_4''\left(\xi_4\right)]-[\omega _{1,5} [\omega _{1,1}
   \omega _{1,3} F_1'\left(\xi_1\right)+\omega _{1,2} \omega _{2,3} F_2'\left(\xi_2\right)]
   F_3'\left(\xi_3\right)\nonumber\\&&+\omega _{2,5} [\omega _{1,1} \omega _{1,4}
   F_1'\left(\xi_1\right)+\omega _{1,2} \omega _{2,4} F_2'\left(\xi_2\right)]
   F_4'\left(\xi_4\right)]{}^2]]\nonumber\\&&/[[\omega _{1,5}
   F_3\left(\xi_3\right)+\omega _{2,5} F_4\left(\xi_4\right)+\omega
   _{3,5}(t)]{}^2],
\end{eqnarray}}where $F_i^{(n)}\left(\xi_i\right)=
\frac{d^nF_i\left(\xi_i\right)}{d\xi_i^n} (i=1, 2)$. All parameters are arbitrary except Eq. (7). In order to discuss the influence of variable coefficients $\gamma$, we take $\gamma=1$ and $\gamma=\cos t$ respectively to show the dynamic properties of solution (8) (see Fig. 2 and Fig. 3).

\includegraphics[scale=0.55,bb=20 320 10 10]{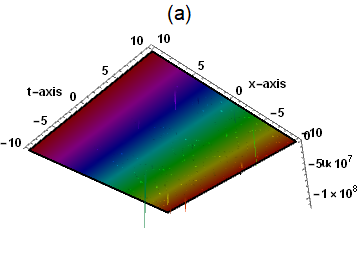}
\includegraphics[scale=0.43,bb=-480 390 10 10]{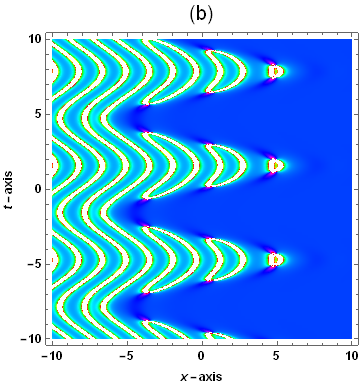}
\vspace{5.8cm}
\begin{tabbing}
\textbf{Fig. 3}. $ \eta_1=\omega _{1,2}=\omega _{2,1}=\omega _{2,3}=\Theta_0=1$, $\eta_2=\omega _{1,1}=\omega _{1,3}=2$, $\omega _{2,2}=-2$,\\ $y=0$, (a) three-dimensional plot,  (b) density plot.
\end{tabbing}

\quad Choosing $F_2(\xi_2)=\cosh \xi_2$ in Eq. (6), we have{\begin{eqnarray} &&\omega _{2,4}=\frac{\omega _{1,4} \omega _{2,3}}{\omega _{1,3}}, \omega _{2,5}=-\frac{\omega _{1,5} \omega _{2,3}^2}{\omega _{2,4}^2}, \omega _{3,1}(t)=\eta _4+[[\omega _{1,1} [-4 \omega _{2,2}^2 \omega _{1,1}^4\nonumber\\ &&+8
   \omega _{1,2} \omega _{2,1} \omega _{2,2} \omega _{1,1}^3+\left(3
   \omega _{1,1}^4+2 \omega _{1,2}^2 \omega _{1,1}^2+3 \omega
   _{1,2}^4\right) \omega _{2,1}^2] \omega _{1,3}^2+[4
   \omega _{1,2}^2 \omega _{2,1}^2 \omega _{1,1}^3\nonumber\\ &&+\left(\omega
   _{1,1}^4-6 \omega _{1,2}^2 \omega _{1,1}^2-3 \omega _{1,2}^4\right)
   \omega _{2,2}^2 \omega _{1,1}+2 \omega _{1,2} (-\omega
   _{1,1}^4+6 \omega _{1,2}^2 \omega _{1,1}^2\nonumber\\ &&+3 \omega _{1,2}^4)
   \omega _{2,1} \omega _{2,2}] \omega _{2,3}^2] \int
   \gamma  \, dt]/[3 \left(\omega _{1,1}^2+\omega
   _{1,2}^2\right){}^2 \left(\omega _{1,1}^2 \omega _{1,3}^2+\omega
   _{1,2}^2 \omega _{2,3}^2\right)],\nonumber\\ && \omega _{3,2}(t)=\eta _3+[[[4 \omega _{1,1}^2 \omega _{2,2}^2 \omega
   _{1,2}^3+\left(-3 \omega _{1,1}^4-6 \omega _{1,2}^2 \omega
   _{1,1}^2+\omega _{1,2}^4\right) \omega _{2,1}^2 \omega _{1,2}\nonumber\\ &&+2
   \omega _{1,1} \left(3 \omega _{1,1}^4+6 \omega _{1,2}^2 \omega
   _{1,1}^2-\omega _{1,2}^4\right) \omega _{2,1} \omega _{2,2}]
   \omega _{1,3}^2+\omega _{1,2} [-4 \omega _{2,1}^2 \omega
   _{1,2}^4\nonumber\\ &&+8 \omega _{1,1} \omega _{2,1} \omega _{2,2} \omega
   _{1,2}^3+\left(3 \omega _{1,1}^4+2 \omega _{1,2}^2 \omega
   _{1,1}^2+3 \omega _{1,2}^4\right) \omega _{2,2}^2] \omega
   _{2,3}^2] \int \gamma  \, dt]\nonumber\\ &&/[3 \left(\omega
   _{1,1}^2+\omega _{1,2}^2\right){}^2 \left(\omega _{1,1}^2 \omega
   _{1,3}^2+\omega _{1,2}^2 \omega _{2,3}^2\right)],\nonumber\\ && \omega _{3,5}(t)=\omega _{1,5} [\frac{\omega _{1,3}^2 \omega _{3,4}(t){}^2}{\omega
   _{1,4}^2}-\omega _{3,3}(t){}^2],\nonumber\\ && \beta=-\frac{\left(\omega _{1,2} \omega _{2,1}-\omega _{1,1} \omega
   _{2,2}\right){}^2 \left(\omega _{1,3}^2-\omega _{2,3}^2\right)
   \gamma }{3 \left(\omega _{1,1}^2+\omega _{1,2}^2\right){}^2
   \left(\omega _{1,1}^2 \omega _{1,3}^2+\omega _{1,2}^2 \omega
   _{2,3}^2\right)},
\end{eqnarray}}where $\eta _3$ and $\eta _4$ are integration constants. Substituting Eq. (6) and Eq. (9) into Eq. (2), the corresponding analytical solution of the second kind for Eq. (1) can be given as follows
{\begin{eqnarray}  &&u=[2 \Theta _0 [\left(\omega _{1,5} F_3\left(\xi_3\right)+\omega
   _{2,5} F_4\left(\xi_4\right)+\omega _{3,5}(t)\right) [\omega _{1,5}
   F_3''\left(\xi_3\right) [\omega _{1,1} \omega _{1,3} F_1'\left(\xi_1\right)\nonumber\\&&+\omega _{1,2}
   \omega _{2,3} F_2'\left(\xi_2\right)]{}^2+\omega _{1,5} F_3'\left(\xi_3\right) [\omega
   _{1,3} F_1''\left(\xi_1\right) \omega _{1,1}^2+\omega _{1,2}^2 \omega _{2,3}
   F_2''\left(\xi_2\right)]\nonumber\\&&+\omega _{2,5} F_4'\left(\xi_4\right) [\omega
   _{1,4} F_1''\left(\xi_1\right) \omega _{1,1}^2+\omega _{1,2}^2 \omega _{2,4}
   F_2''\left(\xi_2\right)]+\omega _{2,5} [\omega _{1,1} \omega
   _{1,4} F_1'\left(\xi_1\right)\nonumber\\&&+\omega _{1,2} \omega _{2,4} F_2'\left(\xi_2\right)]{}^2
   F_4''\left(\xi_4\right)]-[\omega _{1,5} [\omega _{1,1}
   \omega _{1,3} F_1'\left(\xi_1\right)+\omega _{1,2} \omega _{2,3} F_2'\left(\xi_2\right)]
   F_3'\left(\xi_3\right)\nonumber\\&&+\omega _{2,5} [\omega _{1,1} \omega _{1,4}
   F_1'\left(\xi_1\right)+\omega _{1,2} \omega _{2,4} F_2'\left(\xi_2\right)]
   F_4'\left(\xi_4\right)]{}^2]]\nonumber\\&&/[[\omega _{1,5}
   F_3\left(\xi_3\right)+\omega _{2,5} F_4\left(\xi_4\right)+\omega
   _{3,5}(t)]{}^2].
\end{eqnarray}}All parameters are arbitrary except Eq. (9). When $\gamma=1$ and $\gamma=\sin t$,   the dynamic properties of solution (10) are demonstrated in Fig. 4 and Fig. 5.

\includegraphics[scale=0.46,bb=20 370 10 10]{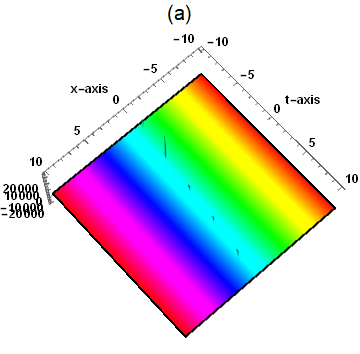}
\includegraphics[scale=0.43,bb=-480 390 10 10]{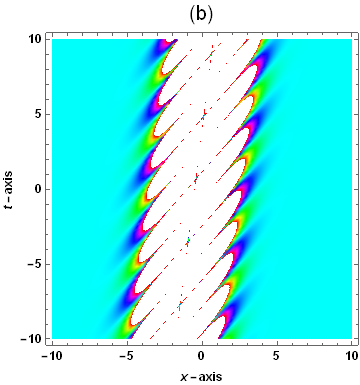}
\vspace{5.8cm}
\begin{tabbing}
\textbf{Fig. 4}. $ \eta_3=\omega _{1,3}=\omega _{2,1}=\omega _{2,3}=\Theta_0=1$, $\eta_4=\omega _{1,1}=\omega _{1,2}=2$, $\omega _{2,2}=-2$,\\ $y=0$, (a) three-dimensional plot,  (b) density plot.
\end{tabbing}

\includegraphics[scale=0.46,bb=20 370 10 10]{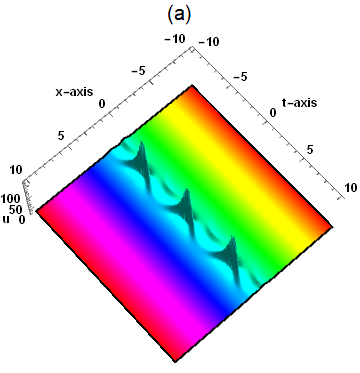}
\includegraphics[scale=0.43,bb=-480 390 10 10]{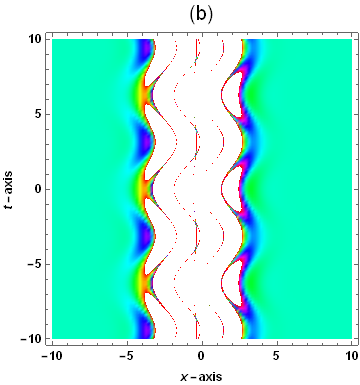}
\vspace{5.8cm}
\begin{tabbing}
\textbf{Fig. 5}. $ \eta_3=\omega _{1,3}=\omega _{2,1}=\omega _{2,3}=\Theta_0=1$, $\eta_4=\omega _{1,1}=\omega _{1,2}=2$, $\omega _{2,2}=-2$,\\ $y=0$, (a) three-dimensional plot,  (b) density plot.\\
\end{tabbing}

\quad Finally, choose a ``3-3-2-1''
variable coefficient neural network model, which denotes that the input layer $l_0=\{x, y, \omega_{l_i}(t)\} (i=1, 2, 3)$ has 3 neurons, the hidden layer $l_1=\{1, 2, 3\}$ has 3 neurons, the hidden layer $l_2=\{4, 5\}$ has 2 neurons and the print layer $\xi$ has 1 neuron. According to the step of vcBNNM, we suppose
{\begin{eqnarray} &&\xi_1= \omega _{3,1}(t)+x \omega _{1,1}+y \omega _{2,1},\nonumber\\
&&\xi_2=\omega _{3,2}(t)+x \omega _{1,2}+y \omega _{2,2},\nonumber\\
&&\xi_3=\omega _{3,3}(t)+x \omega _{1,3}+y \omega _{2,3},\nonumber\\
&&\xi_4=\omega _{1,4} F_1\left(\xi _1\right)+\omega _{2,4}
   F_2\left(\xi _2\right)+\omega _{3,4} F_3\left(\xi
   _3\right)+\omega _{4,4}(t),\nonumber\\
&&\xi_5=\omega _{1,5} F_1\left(\xi _1\right)+\omega _{2,5}
   F_2\left(\xi _2\right)+\omega _{3,5} F_3\left(\xi
   _3\right)+\omega _{4,5}(t),\nonumber\\
&&\xi=\omega _{1,6} F_4\left(\xi _4\right)+\omega _{2,6}
   F_5\left(\xi _5\right)+\omega _{3,6}(t),
\end{eqnarray}}where $F_1(\xi_1)=\cos \xi_1$, $F_2(\xi_2)=\cosh \xi_2$, $F_3(\xi_3)=\cosh \xi_3$, $F_4(\xi_4)=\xi_4^2$, $F_5(\xi_5)=\xi_5^2$, $\omega _{i,j} (i=1, 2; j=1, 2, 3, 4, 5, 6)$ and $\omega _{3,j} (j=4, 5)$ are unknown constants, $\omega _{3,j}(t) (j=1, 2, 3, 6)$ and $\omega _{4,j}(t) (j=4, 5)$ are unknown functions. Substituting Eq. (11) into Eq. (3) by utilizing Mathematica software, we get
{\begin{eqnarray} &&\omega _{2,6}=-\frac{\omega _{1,6} \omega _{2,4}^2}{\omega _{2,5}^2}, \omega _{3,5}=\frac{\omega _{1,5} \omega _{3,4}}{\omega _{1,4}}, \omega _{2,4}=\frac{\omega _{1,4} \omega _{2,5}}{\omega _{1,5}}, \omega _{3,1}(t)=[[\omega _{1,1} [(\omega _{1,1}^2\nonumber\\ &&-3 \omega
   _{1,2}^2) \omega _{1,2}^2+3 \left(\omega _{1,1}^2+\omega
   _{1,2}^2\right) \omega _{1,3}^2] \omega _{2,1}^2+2 \omega
   _{1,2} [-\omega _{1,1}^4+3 \left(\omega _{1,2}^2+\omega
   _{1,3}^2\right) \omega _{1,1}^2\nonumber\\ &&+3 \omega _{1,2}^2 \omega
   _{1,3}^2] \omega _{2,2} \omega _{2,1}+\omega _{1,1}
   [\omega _{1,1}^4-3 \left(\omega _{1,2}^2+\omega
   _{1,3}^2\right) \omega _{1,1}^2-3 \omega _{1,2}^2 \omega
   _{1,3}^2] \omega _{2,2}^2]\nonumber\\ &&* \int \gamma \, dt]/[3
   \left(\omega _{1,1}^2+\omega _{1,2}^2\right){}^2 \omega
   _{1,3}^2]+\eta _6, \omega _{3,2}(t)=[[-\omega _{1,2} [\omega _{1,2}^4-3 \omega _{1,1}^2
   \omega _{1,2}^2\nonumber\\ &&+3 \left(\omega _{1,1}^2+\omega _{1,2}^2\right)
   \omega _{1,3}^2] \omega _{2,1}^2+2 \omega _{1,1} [\omega
   _{1,2}^4-3 \omega _{1,1}^2 \omega _{1,2}^2+3 (\omega
   _{1,1}^2+\omega _{1,2}^2) \omega _{1,3}^2]\nonumber\\ &&* \omega
   _{2,2} \omega _{2,1}+\omega _{1,2} [3 \omega _{1,1}^4-\omega
   _{1,2}^2 \omega _{1,1}^2+3 \left(\omega _{1,1}^2+\omega
   _{1,2}^2\right) \omega _{1,3}^2] \omega _{2,2}^2] \int
   \gamma  \, dt]\nonumber\\ &&/[3 \left(\omega _{1,1}^2+\omega
   _{1,2}^2\right){}^2 \omega _{1,3}^2]+\eta _5,
    \omega _{3,3}(t)=[[8 \omega _{1,1} \omega _{1,2} \omega _{2,1} \omega _{2,2}
   \omega _{1,3}^2+[-3 (\omega _{1,1}^2\nonumber\\ &&+\omega
   _{1,2}^2) \omega _{1,2}^2-\left(\omega _{1,2}^2-3 \omega
   _{1,1}^2\right) \omega _{1,3}^2] \omega _{2,1}^2+[3
   \omega _{1,1}^2 \left(\omega _{1,1}^2+\omega
   _{1,2}^2\right)-(\omega _{1,1}^2\nonumber\\ &&-3 \omega _{1,2}^2)
   \omega _{1,3}^2] \omega _{2,2}^2] \int \gamma  \,
   dt]/[3 \left(\omega _{1,1}^2+\omega _{1,2}^2\right){}^2 \omega
   _{1,3}]+\eta _7,\nonumber\\ && \omega _{2,3}=\frac{\omega _{1,1} \left(\omega _{1,3}^2-\omega _{1,2}^2\right)
   \omega _{2,1}+\omega _{1,2} \left(\omega _{1,1}^2+\omega
   _{1,3}^2\right) \omega _{2,2}}{\left(\omega _{1,1}^2+\omega
   _{1,2}^2\right) \omega _{1,3}},\nonumber\\ && \omega _{3,4}=-\frac{\sqrt{\omega _{1,1}^2+\omega _{1,2}^2} \omega _{1,4}
   \sqrt{\left(\omega _{1,1}^2+\omega _{1,3}^2\right) \omega
   _{1,5}^2+\left(\omega _{1,2}^2-\omega _{1,3}^2\right) \omega
   _{2,5}^2}}{\sqrt{\left(\omega _{1,2}^2-\omega _{1,3}^2\right)
   \left(\omega _{1,1}^2+\omega _{1,3}^2\right) \omega _{1,5}^2}},\nonumber\\ && \beta=\frac{\left(\omega _{1,3} \omega _{2,1}-\omega _{1,1} \omega
   _{2,3}\right){}^2 \gamma }{3 \omega _{1,2}^2 \left(\omega
   _{1,1}^2+\omega _{1,3}^2\right){}^2}, \omega _{3,6}(t)=\omega _{1,6} [\frac{\omega _{1,4}^2 \omega _{4,5}(t){}^2}{\omega
   _{1,5}^2}-\omega _{4,4}(t){}^2],
\end{eqnarray}}where $\eta _5$, $\eta _6$ and $\eta _7$ are integration constants. Substituting Eq. (11) and Eq. (12) into Eq. (2), the corresponding analytical solution for Eq. (1) can be shown as follows
{\begin{eqnarray}  &&u=[2 \Theta _0 [[\omega _{1,6} F_4\left(\xi_4\right)+\omega _{2,6} F_5\left(\xi_5\right)+\omega
   _{3,6}(t)] [\omega _{1,6} F_4''\left(\xi_4\right) [\omega _{1,1} \omega _{1,4} F_1'\left(\xi_1\right)\nonumber\\ &&+\omega _{1,2}
   \omega _{2,4} F_2'\left(\xi_2\right)+\omega _{1,3} \omega _{3,4} F_3'\left(\xi_3\right)]{}^2+\omega
   _{1,6} F_4'\left(\xi_4\right) [\omega _{1,4}
   F_1''\left(\xi_1\right)
   \omega _{1,1}^2\nonumber\\ &&+\omega _{1,2}^2 \omega _{2,4} F_2''\left(\xi_2\right)+\omega _{1,3}^2
   \omega _{3,4} F_3''\left(\xi_3\right)]+\omega _{2,6} F_5'\left(\xi_5\right) [\omega _{1,5} F_1''\left(\xi_1\right) \omega _{1,1}^2\nonumber\\ &&+\omega
   _{1,2}^2 \omega _{2,5} F_2''\left(\xi_2\right)+\omega _{1,3}^2 \omega _{3,5}
   F_3''\left(\xi_3\right)]+\omega _{2,6} [\omega _{1,1} \omega
   _{1,5} F_1'\left(\xi_1\right)+\omega _{1,2} \omega _{2,5} F_2'\left(\xi_2\right)\nonumber\\ &&+\omega _{1,3} \omega
   _{3,5} F_3'\left(\xi_3\right)]{}^2 F_5''\left(\xi_5\right)]-[\omega _{1,6} [\omega _{1,1}
   \omega _{1,4} F_1'\left(\xi_1\right)+\omega _{1,2} \omega _{2,4} F_2'\left(\xi_2\right)\nonumber\\ &&+\omega _{1,3} \omega
   _{3,4} F_3'\left(\xi_3\right)] F_4'\left(\xi_4\right)+\omega
   _{2,6} [\omega _{1,1} \omega _{1,5} F_1'\left(\xi_1\right)+\omega _{1,2} \omega
   _{2,5} F_2'\left(\xi_2\right)\nonumber\\ &&+\omega _{1,3} \omega _{3,5} F_3'\left(\xi_3\right)]
   F_5'\left(\xi_5\right)]{}^2]]/[[\omega _{1,6}
   F_4\left(\xi_4\right)+\omega _{2,6}
   F_5\left(\xi_5\right)+\omega
   _{3,6}(t)]{}^2].
\end{eqnarray}}All parameters are arbitrary except Eq. (12). When $\gamma=1$ and $\gamma=\cos t$,   the dynamic properties of solution (13) are demonstrated in Fig. 6 and Fig. 7.

\includegraphics[scale=0.35,bb=20 380 10 10]{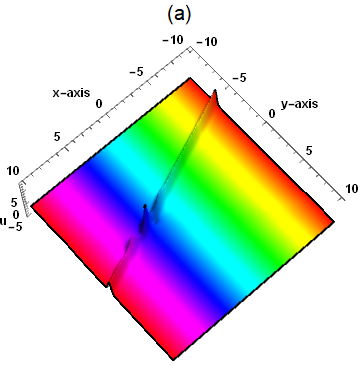}
\includegraphics[scale=0.35,bb=-360 380 10 10]{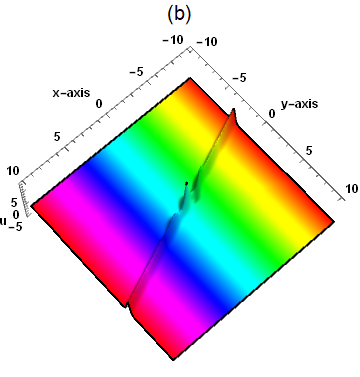}
\includegraphics[scale=0.35,bb=-360 380 10 10]{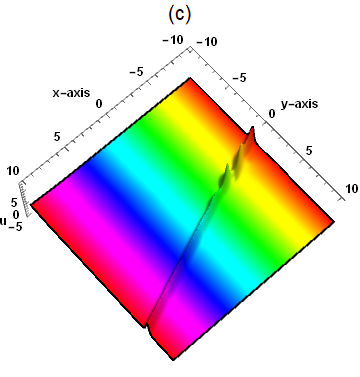}
\includegraphics[scale=0.35,bb=800 780 10 10]{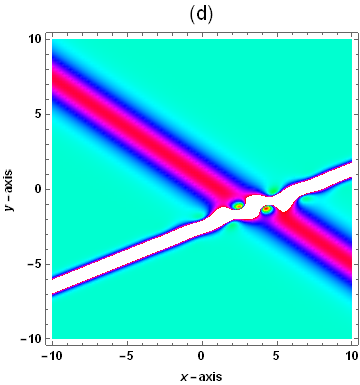}
\includegraphics[scale=0.35,bb=-360 780 10 10]{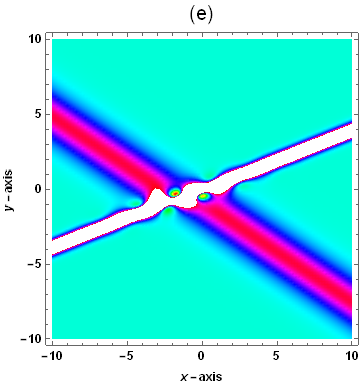}
\includegraphics[scale=0.35,bb=-360 780 10 10]{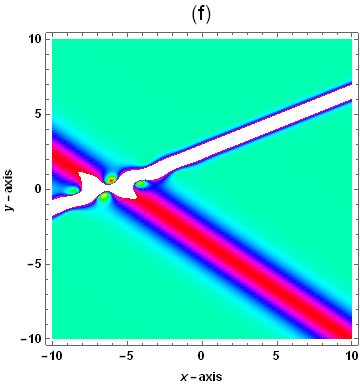}
\vspace{9.5cm}
\begin{tabbing}
\textbf{Fig. 6}. $ \eta_5=\omega _{1,3}=\omega _{1,4}=\omega _{1,5}=\omega _{2,1}=\omega _{2,3}=\Theta_0=1$, $\eta_6=\omega _{1,1}=\omega _{1,2}=2$,\\ $\omega _{2,2}=-3$, $\eta_7=\omega _{2,5}=-1$,  (a)(d) $t=-2$,  (b)(e) $t=0$,  (c)(f) $t=2$.\\
\end{tabbing}

\noindent {\bf 4. Conclusion}\\

\quad The BNNM is a very direct and effective method to obtain the analytical solutions of nonlinear PDEs with constant coefficients, but it is not suitable for solving the analytical solutions of nonlinear PDEs with variable coefficients. Therefore, we propose an improved BNNM named vcBNNM to deal with the problem of analytical solutions of nonlinear PDEs with variable coefficients. To verify the validity of the method, we take the variable coefficient KP equation as an example. By establishing ``3-2-2-1'' and ``3-3-2-1'' models, we successfully obtain a rich analytical solutions of the KP equation with variable coefficients, which contains more arbitrary parameters. Meanwhile, by selecting some special values of the parameters, the dynamics properties of these derived solutions are shown in Fig. 1. Compared with the previous BNNM, we change the parameter $\omega_{l_i}(t)$  in Eq. (5) from constant to function, and the variable $t$ in the input layer $l_0$ also becomes a nonlinear function. In this way, the existence of variable coefficients in the PDEs with variable coefficients can be well handled, and more analytical solutions can be obtained.

\includegraphics[scale=0.46,bb=20 370 10 10]{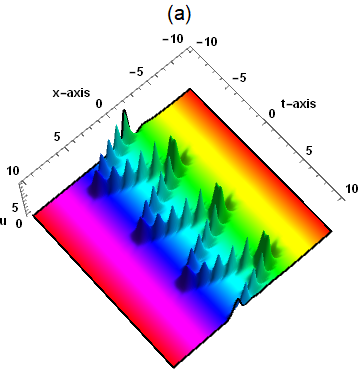}
\includegraphics[scale=0.43,bb=-480 390 10 10]{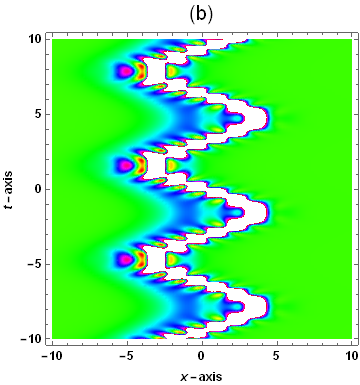}
\vspace{5.8cm}
\begin{tabbing}
\textbf{Fig. 7}. $ \eta_5=\omega _{1,3}=\omega _{1,4}=\omega _{1,5}=\omega _{2,1}=\omega _{2,3}=\Theta_0=1$, $\eta_6=\omega _{1,1}=\omega _{1,2}=2$,\\ $\omega _{2,2}=-3$, $\eta_7=\omega _{2,5}=-1$,  $y=0$, (a) three-dimensional plot,  (b) density plot.\\
\end{tabbing}

\noindent  {\bf Funding:} Project supported by National Natural Science Foundation of
China (Grant No. 12161048), Doctoral Research Foundation of Jiangxi University of Chinese Medicine (Grant No. 2021WBZR007)   and Jiangxi University of Chinese Medicine Science and Technology Innovation Team Development Program (Grant No. CXTD22015).\\

\noindent {\bf Data Availability Statements}\\

Data sharing not applicable to this article as no datasets were generated or analysed during the current study.\\

\noindent {\bf Declaration}\\

\quad {\bf Conflict of interests} The authors declare that there is
no conflict of interests regarding the publication of this article.

\quad {\bf Ethical standard} The authors state that this research
complies with ethical standards. This research does not involve
either human participants or animals.\\

\end{document}